\documentclass[10pt,conference]{IEEEtran}
\IEEEoverridecommandlockouts

\usepackage{cite}
\usepackage{balance}
\usepackage{amsmath,amssymb,amsfonts}
\usepackage{algorithmic}
\usepackage{graphicx}
\usepackage{textcomp}
\usepackage{xcolor}
\usepackage{multirow}
\usepackage{booktabs}
\usepackage{colortbl}
\usepackage{tcolorbox}
\usepackage{pifont}
\usepackage{url}
\def\BibTeX{{\rm B\kern-.05em{\sc i\kern-.025em b}\kern-.08em
    T\kern-.1667em\lower.7ex\hbox{E}\kern-.125emX}}
\begin{document}

\newcommand{\tool}{VulStamp}



\title{
{\tool}: Vulnerability Assessment using Large Language Model}

\author{
\IEEEauthorblockN{Hao Shen$^1$, Ming Hu$^{2*}$\thanks{* Corresponding Authors}, Xiaofei Xie$^{2}$, Jiaye Li$^1$, Mingsong Chen$^{1*}$}
\IEEEauthorblockA{\textit{$^1$MoE Engineering Research Center of SW/HW Co-Design Technology and Application, East China
Normal University, China} \\
\textit{$^2$School of Computing and Information Systems, Singapore Management University, Singapore}
}
}


\maketitle

\begin{abstract}

Although modern vulnerability detection tools enable developers to efficiently identify numerous security flaws, indiscriminate remediation efforts often lead to superfluous development expenses. 
This is particularly true given that a substantial portion of detected vulnerabilities either possess low exploitability or would incur negligible impact in practical operational environments. 
Consequently, vulnerability severity assessment has emerged as a critical component in optimizing software development efficiency.
Existing vulnerability assessment methods typically rely on manually crafted descriptions associated with source code artifacts. 
However, due to variability in description quality and subjectivity in intention interpretation, the performance of these methods is seriously limited.
%
%
To address this issue, this paper introduces VulStamp, a novel intention-guided framework, to facilitate description-free vulnerability assessment.
Specifically, VulStamp adopts static analysis together with Large Language Model (LLM) to extract the intention information of vulnerable code.
Based on the intention information, VulStamp uses a prompt-tuned model for vulnerability assessment.
Furthermore, to mitigate the problem of imbalanced data associated with vulnerability types, VulStamp integrates a Reinforcement Learning (RL)-based prompt-tuning method to train the assessment model. 
%
%
%
%
Extensive experimental results demonstrate that {\tool} outperforms the state-of-the-art baselines by an average of
12.9\%, 102.6\%, 18.3\%, and 54.1\% in terms of AUC, precision, recall and F1-score, respectively.

\end{abstract}


\section{INTRODUCTION}

Software vulnerability~\cite{cao2024snopy, zhao2024coding, zhang2024vuladvisor} refers to the weaknesses or defects in the design, implementation, configuration, operation and other aspects of the software system, which may be maliciously exploited, resulting in serious consequences such as system attacks~\cite{antunes2010vulnerability}, data leakage~\cite{zuo2019does, croft2023data}, business interruption~\cite{dissanayake2022empirical} and so on.
The original CVSS score of CVE-2021-45046 is only 3.7~\cite{balsam2023automated, csw2023ransomware}. 
Ransomware groups have taken advantage of this vulnerability, leading to numerous enterprises being targeted, with their data encrypted and held for ransom.
Upon reevaluation, it was discovered to lead to remote code execution, thereby raising the CVSS score to 9.0~\cite{CVE-2021-45046}.
According to real vulnerability data published on the CVE collected by MegaVul~\cite{megavul} from 2006 to 2023, after statistical analysis, only 12.1\% (820/6,769) are critical-risk vulnerabilities.
Therefore, it is essential to distinguish high-severity, exploitable vulnerabilities from low-risk ones to ensure that remediation efforts are both efficient and cost-effective~\cite{sun2023automatic, xue2025towards}.

Many software vulnerability assessment methods~\cite{han2017learning, sahin2019conceptual, le2019automated, sun2023automatic} based on software vulnerability description have been proposed.
Although its effectiveness has been proven, there is no unified software vulnerability description standard, and the descriptions from different sources and types vary greatly.
In addition, the way software vulnerabilities are exploited and their impact will change with the development of technology and the evolution of attack methods, making it difficult for vulnerability descriptions to keep up with such dynamic changes.
In contrast, the evaluation method based on the characteristics of software code~\cite{xue2025towards, le2022use, hao2023novel} can directly analyze the potential vulnerabilities in the code and avoid the evaluation errors caused by inaccurate, vague or incomplete descriptions.
However,  existing methods still face three challenges, i.e., \ding{182} noise code pollution, \ding{183} incomplete intention analysis, and \ding{184} limited critical-risk vulnerability attention, which is detailed as follows:


\textbf{Challenge 1: Incomplete intention analysis.}
Severity assessment typically requires an understanding of vulnerability intention, which refers to the potential behavioral objectives an attacker can achieve by exploiting a vulnerability under specific conditions~\cite{dang2015comparing}, such as memory corruption~\cite{iannone2022secret}, privilege escalation~\cite{davis2004processes}, and information disclosure~\cite{shahzad2012large, sabottke2015vulnerability}. 
However, existing methods mainly rely on learning vulnerability patterns directly from source code, without information about the intentions of vulnerabilities. 
As a result, these models lack the ability to accurately identify the real trigger conditions and impact pathways of vulnerabilities, leading to limited effectiveness in severity assessment. 
Although some methods~\cite{xue2025towards} try to use code descriptions to improve the performance of the assessment, the quality and consistency of these descriptions can vary significantly due to differences in the expertise of auditors and subjective interpretations. 
As a result, the performance of such methods is still severely constrained.


\textbf{Challenge 2: Noise code pollution.}
Typically, vulnerabilities usually appear only in local regions of the code.
However, most of the existing methods rely on coarse-grained representations at the function level for vulnerability assessment~\cite{xue2025towards, le2019automated, le2022use}, which contain a large number of irrelevant code elements (such as variable declarations, irrelevant control flow, logging statements, etc.), forming a serious semantic noise.
Even advanced language models such as ChatGPT~\cite{chen2025chatgpt} are contaminated by semantic noise when faced with large amounts of contextual code.

\textbf{Challenge 3: Limited critical-risk vulnerability attention.}
Most of the existing methods are based on the deep learning model, whose performance are seriously limited by the quality of training data.
However, in the actual software vulnerability datasets, the number of critical-risk samples is often far less than that of medium- and low-risk vulnerabilities~\cite{megavul, yan2024stealing}. 
This seriously unbalanced distribution forms a typical long-tail problem~\cite{wen2024livable} at the data level.
This deviation makes the model prone to misjudge the severity of security-critical vulnerabilities, resulting in serious security risks.


\textbf{Insight.} 
Intuitively, to address the above challenges, since Large Language Models (LLMs) exhibit the powerful capability of language understanding and reasoning, they are promising in extracting the intention of vulnerable code.
To address the noise code pollution problem, intuitively, we can adopt static analysis technologies to filter out irrelevant code segments and guide the model to focus on the key code that is most relevant to vulnerability semantics, reducing distraction from unrelated logic.
To address the problem of unbalanced data, we can optimize the training strategy to place greater emphasis on high-risk vulnerabilities.

\textbf{Our work.}
Inspired by the above insights, we present a novel vulnerability assessment framework, named {\tool}, which integrates the vulnerability code syntax analysis with the intention analysis by LLM to improve vulnerability assessment.
Specifically, {\tool} consists of three main modules.
Firstly, the program dependence graph of the code is constructed, and the code parts related to the intention of the vulnerability are preserved by slicing forward and backward according to the vulnerability interest point.
Next, the LLM is used to generate the exploitability, impact, and scope of the vulnerability from the code to report the attack intention.
Finally, the reward function for the vulnerability to the serious risk concerned was constructed to enhance the attention and consistency of the representation of the features of the minority class.
We developed a prototype system called {\tool} and constructed a dataset of 6,769 real software vulnerabilities that comply with the CVSS 3.0 standard.
Experimental results show that compared with the state-of-the-art vulnerability assessment method~\cite{xue2025towards}, {\tool} improves AUC, precision, recall, and F1-score by 7.8\%, 39.4\%, 8.4\% and 21.6\%, respectively.
This paper makes the following \textbf{contributions}:
\begin{itemize}
    \setlength{\leftmargin}{\parindent} 
    \item We propose {\tool}, a method that uses code simplification grammar rules to extract code intention from vulnerability samples containing a large amount of semantic noise for effective vulnerability evaluation.
    \item We propose a novel LLM-based method to extract vulnerability intention reports, which enhances the ability of the model to infer vulnerability intentions.
    \item We construct the reward function for the vulnerability of the serious risk of concern and enhance the distinguishability of the representation of the vulnerability feature.
    \item We evaluate {\tool} on the constructed dataset, and the results demonstrate the effectiveness of {\tool} in software vulnerability assessment.

\end{itemize}

\section{BACKGROUND AND MOTIVATION}

\begin{figure}[htpb]
\centering	
    \includegraphics[width=0.45\textwidth]{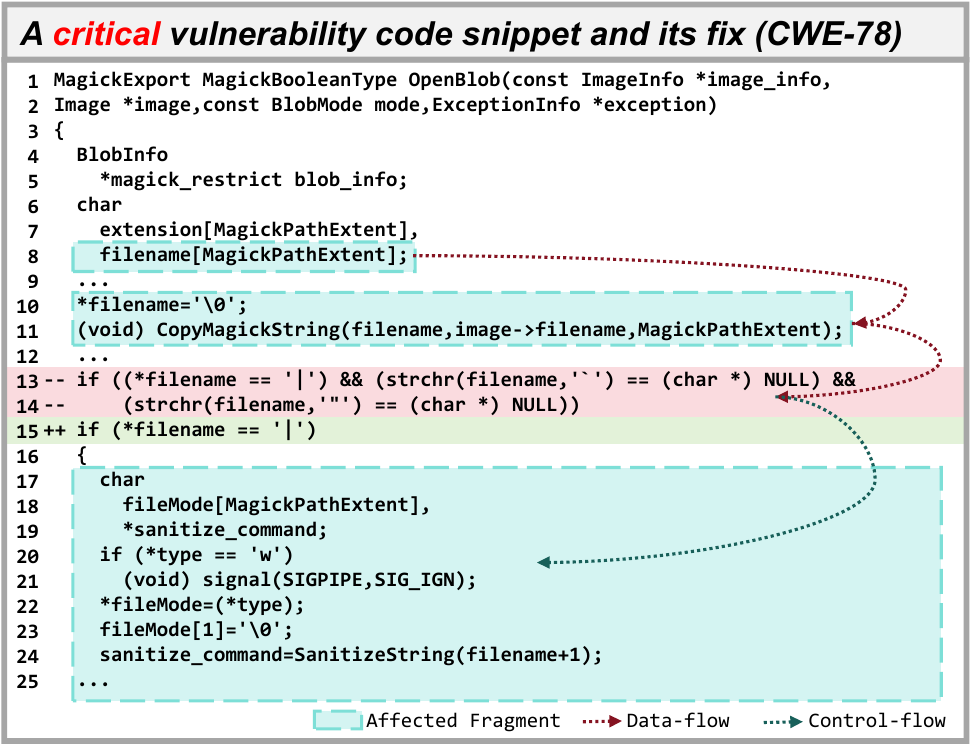}
    \vspace{-0.1in}
    \caption{An example of code illustrating a command injection vulnerability within the ImageMagick project that was inaccurately rated as low-risk by the SVACL. In the depiction, segments of code that were updated before and after the patch are highlighted using red and green, respectively.}
    \label{fig:movitation}
    \vspace{-0.12in}
\end{figure}

\subsection{Background}

CVSS (Common Vulnerability Scoring System) \cite{cvss1,cvss,cvss3,cvssicse} is a standardized scoring system used to measure and evaluate the severity of vulnerabilities.
It aims to provide a unified, objective, and quantifiable approach to the field of cybersecurity to measure the potential risks of vulnerabilities, helping organizations and enterprises more effectively identify, prioritize, and fix vulnerabilities and rationally allocate resources to deal with cybersecurity threats.

Previous studies mainly used CVSS 2.0 \cite{cvss1}, which introduced basic evaluation indicators and life cycle evaluation indicators, and was widely applied in software suppliers and enterprises.
The scoring system of CVSS 2.0 requires users to have an overly detailed understanding of the exact impact of vulnerabilities, and the scoring range is not comprehensive enough when faced with some new types of attack or complex vulnerability scenarios.
CVSS 3.0 \cite{cvss3,cvssicse} introduces the concept of the ``scope'' to distinguish whether the affected components and the vulnerable components are the same.
The weight distribution of the basic indicators has also been adjusted, the granularity of the indicators has been refined, and a new ``severe'' level has been added to make it applicable to a wider range of scenarios.
Therefore, recent vulnerability reports mainly use the CVSS 3.0 score, so we also adopt CVSS 3.0 as our experimental standard.
 
\subsection{A Motivating Example}
\label{sec:movitation_example}

Figure~\ref{fig:movitation} shows an example of fixing a possible command injection vulnerability (CVE-2023-34152)~\cite{CVE-2023-34152} in ImageMagick.
We emphasize the importance of directly inferring the intent of software vulnerabilities by combining vulnerability intent reporting and reducing vulnerability noise.
In the vulnerable function, when \texttt{filename} starts with \texttt{|}, it's a shell pipe command, and OpenBlob executes it via \texttt{popen()}.
The conditions are: No backticks \texttt{(')}, no double quotes \texttt{(")}.
It looks like a security check, but it is actually a wrong "blacklist" way of defense.
The original restriction would reject legitimate pipeline commands that contain special characters such as \texttt{|sh -c "convert input.png output.jpg"} or command templates that contain \texttt{' ' '}.
This strict restriction prevents some advanced usages from working.
And this may prevent some users from doing things like:
\texttt{convert 'image.png' '|sh -c "process  $\backslash$ "arg with spaces  $\backslash$ ""'}
The double quotes are necessary, but the logic of the vulnerable code rejects them.
However, existing severity assessment methods depend on the context and focus only on the local part of the code, without specifying where the filename comes from and whether there are additional sandboxes/permission restrictions.
Then existing approaches may default to "narrow attack surface" and thus conservatively consider this as ``medium-risk'' or ``low-risk''.
With this example, we demonstrate the following insights:

\textbf{Mitigating noise to more fully track code intention.}
By analyzing the above CWE-78 vulnerability code, we can see that over-reliance on the local context can lead to noise pollution of code understanding and cause wrong evaluation.
Existing vulnerability assessment techniques, such as SVACL~\cite{xue2025towards} and GraphEval~\cite{hao2023novel}, understand the patterns of vulnerability functions by parsing the code into vulnerability attribute graphs or combining continuous learning, but ignore that there is much noise in the vulnerability functions, which greatly affects the model's understanding of the intent of the code.
For software vulnerabilities, making appropriate rules to alleviate noise and trace the intent above and below the code can help to more comprehensively assess the risk of the vulnerability.

\textbf{Extracting vulnerability intentions more semantically to exploit the evaluation.}
In the \texttt{filename} check above, the value of the filename is directly related to the feasibility of access.
However, not every model can immediately recognize that \texttt{popen}  executes user-controlled input as command injection. They pay more attention to syntax and the execution process, ignoring how the attacker's input affects system command execution.
Some models even misinterpret command injection as a failed file opening or external command execution, thus underestimating the risk.
The vulnerability report can automatically ``complete'' the attacker's perspective and understand the intent of the vulnerability.
It can summarize the vulnerability intention more semantically, identify the exploitation conditions of the business layer and the consequences of attacks, and assist in automated security assessment.

\section{METHODOLOGY}

\subsection{Overview of {\tool}}
\begin{figure}[h]
\centering	
    \includegraphics[width=0.499\textwidth]{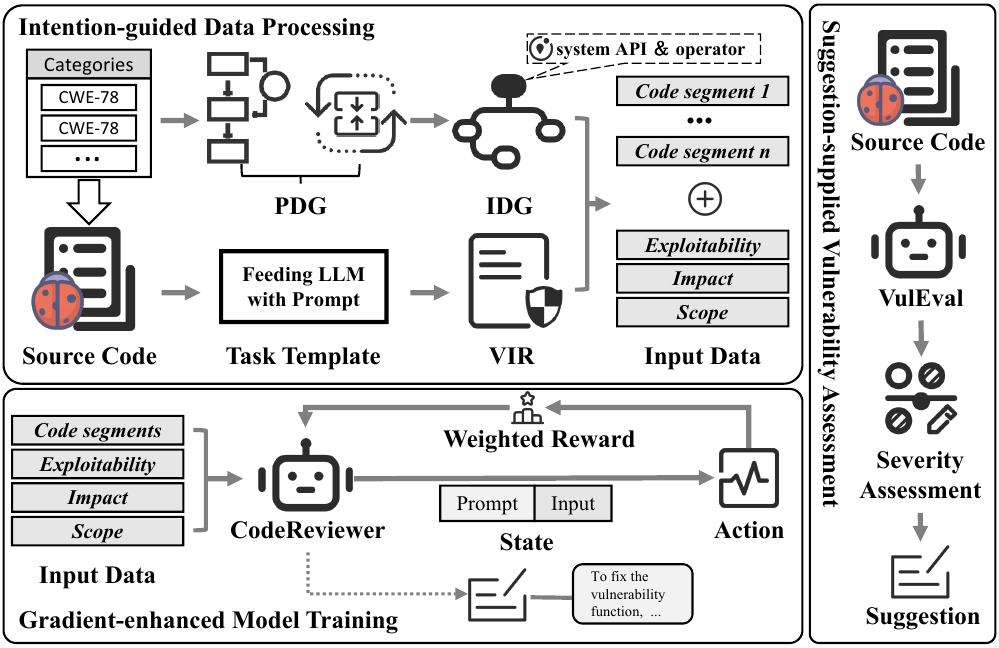}
    \vspace{-0.15in}
    \caption{Framework and workflow of {\tool}.}
    \label{fig:fw}
    \vspace{-0.1in}
\end{figure}

Figure~\ref{fig:fw} illustrates the overall framework of {\tool}, which mainly consists of the following three modules:
\ding{182} \textbf{Intention extraction} adopts static analysis to denoise the vulnerability code into multiple key code segments and LLM to generate the Vulnerability Intention Report (VIR).
{\tool} combines code segments together with VIRs as intention data used for model training or vulnerability assessment.
\ding{183} \textbf{Model training} captures intention information from training data and adopts an RL-based prompt-tuning strategy to train the deep learning model used for vulnerability assessment.
\ding{184} \textbf{Vulnerability assessment} adopts the extracted intention data as input to the trained deep learning model for vulnerability assessment and repair suggestion.
The details of {\tool} are described below.

\subsection{Intention Extraction}
The purpose of this part is to process the target vulnerability function to extract the vulnerability intention representation.
As shown in Section~\ref{sec:movitation_example}, injecting an entire vulnerability function into a model inevitably introduces code noise unrelated to the intention of the vulnerability, and these models suffer from redundant information when trying to understand the intention of the code.
Inspired by the achievements of vulnerability fixing commits and LLMs~\cite{yin2024multitask, wang2024contracttinker, zhang2024vuladvisor}, we propose a strategy based on intention-guided intent denoising and intent analysis, which performs code slicing according to code intent trigger points and utilizes the extensive code knowledge of LLMs to bridge the intrinsic analysis of vulnerability modules.

\textit{\bf Vulnerability Intent Tracking.}
To avoid the noise of redundant information when trying to understand the intent of the code, we introduce a set of code denoising rules, select the system API calls and operators as the point of interest according to the characteristics analyzed in Section~\ref{sec:movitation_example}, and perform program slicing on the vulnerability function.
First, we use the mature static analysis tool Joern\footnote{https://github.com/joernio/joern} to parse the source code of the program and get the program dependence graph.
Then, we also obtain the call graph of the code through the pycallgraph library\footnote{https://github.com/gak/pycallgraph}, which is used to supplement the semantic information, such as call relations and return values, in the PDG for a comprehensive control data flow analysis.
We use system APIs~\cite{cheng2021deepwukong} and operators that are widely used but misused by applications as points of interest.
It also divides the operators into four categories: arithmetic operators, bitwise operators, compound assignment expressions, and increment/decrement expressions.
Next, we slice forward~\cite{yadavally2024learning} and backward~\cite{thome2018integrated} depending on which node of the program we are interested in.
The purpose of forward slicing~\cite{yadavally2024learning} is to trace the construction process of the source data, so that the intent of the code to prepare the data before calling the point of interest is clear.
Such as where the arguments to a system API come from and whether they are properly validated or initialized.
Therefore, we will only focus on the path from which the input data comes, that is, the data flow.
The purpose of backward slicing~\cite{thome2018integrated} is to fully track where the output data goes and how it affects the subsequent code, helping us to understand how the results of key operations affect the subsequent behavior of the program and understand the intentions of the code to process the output data of  POI.
For example, does a change to a variable cause a subsequent condition to fail or trigger an unsafe operation?
So we focus on both data flow and control flow.
Finally, through forward and backward slicing, we can clarify the intention relationship between the input and output of the interest point, eliminate redundant information irrelevant to the interest point, and obtain the Intention Dependence Graph (IDG).
The code retained in the intent dependency graph enables us to identify the potential vulnerability points more accurately, which helps us better understand the design intention of the code.
As shown in Figure~\ref{fig:movitation}, the affected fragment is the intention dependency graph retained after slicing.

\textit{\bf Program Intent Extraction.}
To understand the impact of the vulnerable code for better evaluation, we also focus on the three Vulnerability Intent Reports (VIRs) of the function, which are exploitability, impact, and scope.
Exploitability refers to the ease with which a vulnerability can be exploited by an attacker. It evaluates the conditions required by an attacker to exploit the vulnerability, the level of skill required, the information and resources required, etc.
To structurally express the exploitability of a vulnerable function, we introduce the EXP (condition, way) notation, where condition denotes the condition that needs to be satisfied before a vulnerability can be exploited, and way denotes how the attacker can access the vulnerable code.
Impact refers to the result of the system or application after the vulnerability is exploited. This includes data tampering, information leakage, denial of service, and other aspects.
Clarifying the impact of a vulnerability helps us determine the priority of the vulnerability and how to evaluate the effect of the fix after it has been fixed.
We structurally denote impact as IMP (impact), where impact represents the possible consequence if a vulnerability is successfully exploited.
The scope refers to the boundary of the impact of the vulnerability, that is, the spread and impact range of the vulnerability in the system or network.
This involves analyzing whether the vulnerability can be confined to a specific module or component or whether it may spread to the whole system or even to other connected systems.
Similarly, we denote the scope as SCO (scope), where scope denotes the scope affected by the vulnerability.
Knowing the scope helps us plan our strategy to fix the vulnerability.

\begin{figure}[htpb]
\centering	
    \includegraphics[width=0.45\textwidth]{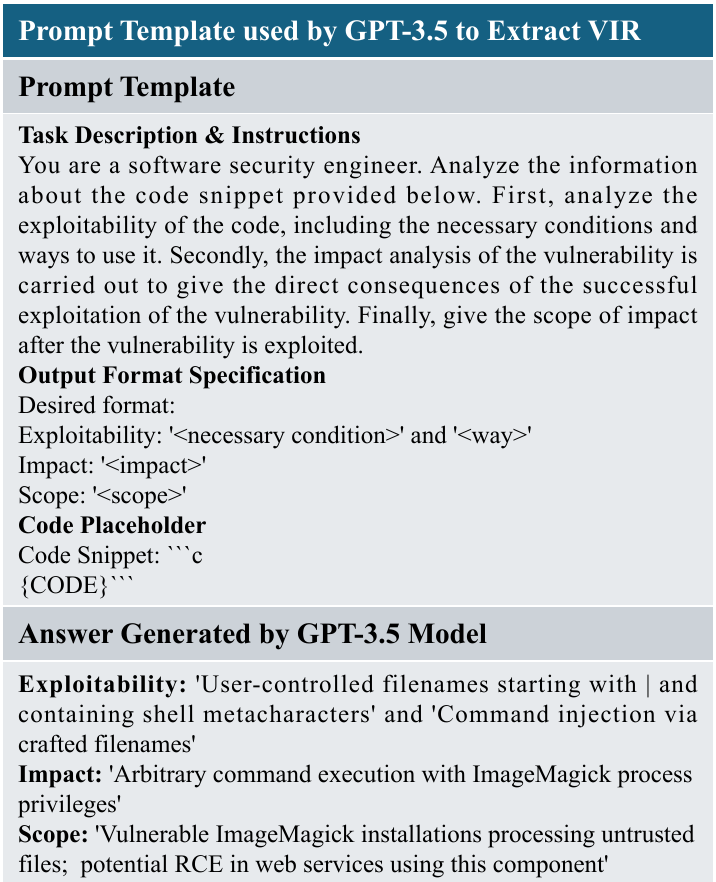}
    \vspace{-0.1in}
    \caption{The prompt template used for extracting VIRs.}
    \label{fig:prompt}
    \vspace{-0.15in}
\end{figure}

As shown in Figure~\ref{fig:prompt}, we build a prompt template to handle the given vulnerability function.
Then, the vulnerability function and prompt template are entered into the LLM to obtain the vulnerability intention.
According to existing research and OpenAI practice, the prompt template designed by us consists mainly of three parts: \text{Task Description \& Instructions}, Output Format Specification, and Code Placeholder.
Task Description \& Instructions Designates the user as a software security engineer, emphasizes his/her responsibilities and professional background, and ensures that the user performs the analysis in the right context.
The Output Format Specification specifies three parts of our output: Exploitability, Impact, and Scope, and provides a concise way to describe each.
A uniform output format facilitates quick access to key information for easy integration into reports or databases, as well as quick understanding of analysis results by non-technical stakeholders.
The Code Placeholder provides a place for the code snippet to be analyzed, so that the analysis task can focus on the specific code and the analysis can be targeted and accurate.

We apply this template for the code in~Figure~\ref{fig:movitation}.
Then, the GPT-3.5 model is used to generate the corresponding vulnerability intent report, as shown in~Figure~\ref{fig:prompt}.
By analyzing the code snippets and evaluating their exploitability, impact, and scope, it shows that the GPT model can deeply understand and analyze vulnerability functions.

\subsection{Model Training}
In the vulnerability assessment task, we treat the model that performs the vulnerability severity assessment as an agent.
First, the agent aims to learn how to accurately assess the severity of code vulnerabilities so that it can quickly identify and prioritize critical-risk vulnerabilities in real-world applications.
Secondly, the agent perceives the state of the environment through the input IDG and VIR. 
After processing, the input data is encoded into representations that can be processed by the model.
Then, based on the perception of the state of the environment, the agent uses its policy to assess the severity of the vulnerability.
Finally, the agent outputs the severity of the evaluation in the environment and receives the reward signal from the environment as feedback to adjust its policy to take better actions in the future.

\textit{\bf Prompt Tuning Model.}
Prompt tuning guides a pre-trained model to produce output in a specific format or content by adding learnable hints to the input text, rather than directly fine-tuning the model's parameters.
Depending on the type of prompt word, there are three ways: hard prompts, soft prompts, and hybrid prompts~\cite{cai2025adapting}.

Hard prompts are the direct insertion of predefined, fixed natural language templates or words into the input text, which are not adjusted during training.
Soft hints refer to the insertion of learnable vectors or tokens into the input text, which are learned to be optimized along with the model parameters during training.
However, hybrid prompts combine the advantages of hard and soft prompts, containing fixed natural language templates (i.e., hard prompts) and learnable vectors or tags (i.e., soft prompts).
The {\tool} prompt template is defined as follows:
\begin{equation}
\begin{aligned}
f_{hybrid} = &\text{$The$ $code$ $snippet$ $:$ $[X]$ $The$ $vulnerability$ } \\ 
             &\text{ $analysis$ $:$ $[Y]$ $[SOFT]$ $[Z]$},
\end{aligned}
\end{equation}
where ``The code snippet:'' and ``The vulnerability analysis:'' are hard prompt parts that explicitly inform the model about the type of input text.
``Classify the severity:'' is the soft prompt part, denoted $[soft]$, which can be replaced by ``identify the severity of this vulnerability:'', etc., to guide the model in displaying the severity classification results through the learnable vector representation.
It combines the interpretability of hard prompts and the flexibility of soft prompts, which can not only guide the model to output results in specific formats, but also optimize the prompt effect through learning.


\textit{\bf Weighted Reward Function.}
A model's prediction confidence (i.e., the maximum predicted probability) measures how ``confident'' the model is in its current prediction.
The higher the confidence, the higher the reward can be.
If the evaluation is correct, the reward is the positive value of the confidence.
If the evaluation is wrong, the reward is negative and is inversely proportional to confidence.

Furthermore, misclassifying ``high risk'' as ``low risk'' can be more serious than misclassifying ``low risk'' as ``high risk'' in vulnerability assessment.
To make the model pay more attention to predicting high-risk categories, we introduce a weighted reward function to set different reward weights for different severity categories.
Based on the severity corresponding to the scores designed by CVSS 3.0, we take the median severity score of each severity to assign weights to the vulnerability functions of different degrees. 
The formula is as follows:
\begin{equation}
r = 
\begin{cases} 
    w \cdot p, & \text{if } \hat{y} = y \\
    -w \cdot p, & \text{if } \hat{y} \neq y 
\end{cases},
\end{equation}
where $w$ is the class weight, the high-risk class has a higher weight, and $p$ is the probability that the model predicts the current class $\hat{y}$, $y$ is the true class.
This reward mechanism penalizes the model more for incorrect predictions with high confidence and rewards it more for correct predictions with high confidence.

\textit{\bf Reinforcement Learning.}
We introduce the update of the reward baseline, which is used to reduce the variance of the reward and make the training process of Reinforcement Learning (RL) more stable. 
The formula is as follows:
\begin{equation}
b_{\text{new}} = \alpha \cdot b_{\text{old}} + (1 - \alpha) \cdot \bar{r} ,
\end{equation}
where $\alpha$ is the momentum attenuation coefficient and $\bar{r}$ is the average reward of the current batch.
The reward baseline helps the model learn more efficiently by smoothing out the changes in the reward signal.

The loss of the policy gradient is used to optimize the model policy so that the actions taken by the model in a given state lead to higher long-term rewards.
It optimizes the model's policy by maximizing the expected reward, which is computed as follows:
\begin{equation}
    L_{PG} = - \mathbb{E}_{s \sim D, a \sim \pi_{\theta}} \left[ \log \pi_{\theta}(a \mid s) \cdot (R(s, a) - b) \right] ,
\end{equation}
where $\pi_{\theta}(a \mid s)$ denotes the probability function, $D$ is the state distribution, $R(s, a)$ is the reward obtained by taking action $a$ in state $s$, and $b$ is the reward baseline.

\textit{\bf Multi-task Learning.}
Previous studies have shown that multi-task learning is effective in improving performance on learning-based tasks such as code completion, code generation, and code understanding.
We also introduce a multi-task learning framework to improve the model's understanding of vulnerability patterns for better evaluation.
Specifically, according to the vulnerability dependency graph and vulnerability intention report, {\tool} further provides repair suggestions for the vulnerability function after evaluating the vulnerability.
That is, {\tool} is trained on two tasks, vulnerability assessment and vulnerability fix proposal generation, respectively.
The total loss is the sum of the losses for each task and is calculated as follows:
\begin{equation}
    L_{total} = L_{assessment} +  L_{suggestion} +  \lambda  L_{PG} ,
\end{equation}
where $L_{assessment}$ is the vulnerability assessment loss,  $L_{PG}$ is the policy gradient loss, $L_{suggestion}$ is the vulnerability repair suggestion generation loss, and $\lambda$ is the weight coefficient of the policy gradient loss.

\subsection{\bf Vulnerability Assessment}
\label{sec:fw_va}
After model training, {\tool} can be efficiently deployed on consumer-grade graphics cards to evaluate vulnerability functions and provide repair suggestions.
Specifically, similar to the intention-guided data processing part, the intention dependency graph is extracted through program analysis, and the LLM is used to extract vulnerability intention reports.
The IDG and the VIR are then concatenated together.
Finally, the vulnerability information is fed to {\tool} to asses vulnerability and give repair suggestions.

\section{EXPERIMENTS}

\subsection{Baselines}
\label{baselines}
To demonstrate the effectiveness of our {\tool} method, we compare it with three major types of baseline methods.

\textbf{Supervision-based Methods.}
We considered three of the latest supervision-based methods in the field of software vulnerability assessment, i.e., CWM (Character-word Model)~\cite{le2019automated}, Fun (Function-level Support Vector Analysis)~\cite{le2022use}, and SVACL~\cite{xue2025towards}. Among them, CWM and Fun both take the vulnerability code as input and give the evaluation results, while SVACL additionally considers the description information of vulnerability functions. In our experiment, we analyze the performance of $\text{CWM}_{SVM}$, $\text{CWM}_{XGB}$, $\text{Fun}_{RF}$, and $\text{Fun}_{LGBM}$ based on the different classifiers utilized.

\textbf{Pretrained Model-based Methods.}
We examined six approaches using general-purpose pretrained code models, which are widely used for various code-related downstream tasks.
Specifically, we considered three encoder-based models (i.e., CodeBERT~\cite{feng2020codebert}, GraphCodeBERT~\cite{guographcodebert}, and RoBERTa~\cite{liu2021robustly})  and three encoder-decoder based models (i.e., CodeReviewer~\cite{li2022automating}, Unixcoder~\cite{guo2022unixcoder}, and CodeT5~\cite{wang2021codet5}). All these models take the same inputs and outputs as CWB and Fun.

\textbf{LLM-based Methods.}
We investigated eight mainstream large code language models for vulnerability assessment performance comparison, including Qwen2.5-Coder-7b, Star-Coder2-7b, Deepseek-Coder-7b, CodeLlama-7b, GPT-3.5-turbo, GPT-4-turbo, DeepSeek-V3, and DeepSeek-R1.


\subsection{Datasets for Evaluation}

To evaluate the vulnerability assessment performance of {\tool}, we adopted MegaVul \cite{megavul}, a dataset to evaluate real vulnerabilities in C/C++, including 17,380 vulnerabilities sourced from 992 open-source repositories over the past two decades, covering 169 unique vulnerability types.
The MegaVul dataset offers a broader time span and a more diverse array of both vulnerability sources and types when compared to the Devign~\cite{devign} and Big-Vul~\cite{bigvul} datasets. Moreover, it allows for updates whenever new vulnerabilities are identified.
In MegaVul, each vulnerability instance is clearly labeled and comes with an in-depth description that covers the type of vulnerability, its severity, as well as additional information, offering a dependable foundation for model training and evaluation.
During our experiment, we chose vulnerability functions adhering to the CVSS 3.0 standard, and assigned severity levels to the vulnerabilities: 0 indicating low risk, 1 indicating medium risk, 2 indicating high risk, and 3 indicating critical risk.

\textbf{Suggestion Collection.}
To provide repair suggestions after vulnerability assessment, we prepared vulnerability repair suggestions according to the suggestion extraction method, i.e.,   VulAdvisor~\cite{zhang2024vuladvisor}.
%
We utilized OpenAI's publicly available API, ``gpt-3.5-turbo'', known for its efficiency in producing responses.
When errors occur in the generation process, we dismiss these errors and produce suitable alternatives.
Furthermore, we removed the generated duplicate or summary code comments and regenerated the suggestions.


\textbf{Dataset Division.}
Typically,  LLM-based training involves dividing a dataset into three sets: the training set, the validation set, and the test set.
As shown in Table~\ref{tab:timeinfo},  SVACL~\cite{xue2025towards} randomly divides MegaVul into three sets, where all samples were collected between 2014 and 2025. In this case, all three sets have similar time intervals for collecting samples, neglecting the impact of the disclosure time of vulnerability samples.
Obviously, this division does not reflect the practical vulnerability assessment scenario, where models are trained first and then used to assess unexplored vulnerabilities. In other words, the partition of the dataset by SVACL~\cite{xue2025towards} is unfair for comparing the performance of vulnerability assessment, providing models with a ``prophet'' advantage in the test set.


\begin{table}[h]
\vspace{-0.1in}
  \centering
  \caption{Comparison of Dataset Partitioning Schemes between SVACL and VulStamp}
  \footnotesize
    \begin{tabular}{ccccc}
    \toprule
    \textbf{Method} & \textbf{Time Point} & \textbf{Training} & \textbf{Validation} & \textbf{Test} \\
    \midrule
    \multirow{1.5}[2]{*}{SVACL} & Start & 20140516 & 20141110 & 20141110 \\
          & End   & 20250311 & 20250213 & 20250122 \\
 \midrule
    \multirow{1.5}[2]{*}{Ours} & Start & 20140516 & 20220818 & 20220818 \\
          & End   & 20220817 & 20250226 & 20250311 \\
    \bottomrule
    \end{tabular}%
  \label{tab:timeinfo}%
  \vspace{-0.05in}
\end{table}%

To address this time travel issue,  we crawled the published data for each sample from the CVE website$\footnote{https://cve.mitre.org/}$, and divided the MegaVul dataset according to the chronological order of samples. As shown in Table~\ref{tab:timeinfo}, we assumed that all training data was collected before the time of occurrence of the validation and test samples.
Moreover, we gathered associated scores from the CVE website to complete the vulnerability information for samples that had not been assessed using CVSS 3.0 in the dataset. Note that we use our dataset partitioning scheme throughout the paper.

\subsection{Experimental Settings}
\label{sec：exp_set}
All experiments were performed on a Linux server equipped with Intel(R) Core(TM) i9-12900k and 24GB of NVIDIA GeForce RTX 4090 GPU.

\textbf{Implementation Details.}
We utilized the publicly available source code and hyperparameters originally provided by the authors for the CWB, Fun and SVACL methods.
For pre-trained model-based methods, we downloaded the available models from HuggingFace and then evaluated them on our server.
For GPT-3.5-turbo and DeepSeek-R1, we used the common API ``gpt-3.5-turbo'' of OpenAI and the common API ``deepseek-reasoner'' of DeepSeek for vulnerability assessment, respectively.
Our model of {\tool} was constructed using the PyTorch library and the Transformers library. Additionally, we used the ``gpt-3.5-turbo'' API to extract code intentions, and optimized the fine-tuning of the pre-trained model CodeReviewer using the Adam optimizer at a learning rate of 5e-5.
We used a mature code parser, Joern, to generate the code PDGs.
For efficient model fine-tuning, we utilized the OpenPrompt library, employing a batch size of 16 and executing 100 training iterations.
The coefficient of reinforcement learning loss is configured at 0.01, while the momentum decay coefficient for the reward baseline is adjusted to 0.7. The impact of these choices will be discussed in Section~\ref{sec:rq4}.

\textbf{Evaluation Metrics.}
We conducted the performance evaluation using four indicators, i.e., AUC, recall, precision, and F1-score.
The AUC, or area under the ROC curve, can be computed using a one-to-many approach, specifically One-vs-Rest (OvR).
The recall metric reflects the proportion of vulnerabilities at a given severity level that are accurately evaluated as that same severity level.
The precision metric indicates the proportion of evaluated vulnerabilities corresponding to that severity level.
The F1-score, which indicates the joint efficacy of precision and recall at a certain severity level, is calculated using the formula $2 \times \frac{Recall \times Precision}{Recall + Precision}$.
Note that the higher the metric values, the better performance we can achieve.

\section{EXPERIMENTAL RESULTS}

This section presents various experiments to evaluate the effectiveness of {\tool}.  We compared {\tool} with state-of-the-art (SOTA) vulnerability assessment methods, aiming to answer the following four Research Questions (RQs).

\textbf{RQ1:} How is the {\bf superiority} of {\tool} compared with SOTA vulnerability assessment methods? 

\textbf{RQ2:} How is the {\bf effectiveness} of the proposed key components on the performance of {\tool}?

\textbf{RQ3:} How is the {\bf generalization} ability of {\tool} when applied to other pretrained models? 

\textbf{RQ4:} How is the {\bf impact} of  hyper-parameters on the performance of {\tool}?


\subsection{Performance Evaluation (RQ1)}

We compared {\tool} with the SOTA software vulnerability assessment methods described in Section \ref{baselines}. \tablename~\ref{tab:rq1_1} presents the comparison results, where the best results are highlighted in bold and the second-best results are underlined.  

\begin{table}[h]
  \vspace{-0.15in}
  \centering
  \caption{Performance comparison between  {\tool} and  baselines.}
  \vspace{-0.05in}
  \footnotesize
    \begin{tabular}{clcccc}
    \toprule
    \rowcolor[rgb]{ .749,  .749,  .749}  \textbf{Type} & \textbf{Method}    & \textbf{AUC} & \textbf{Precision} & \textbf{Recall} & \textbf{F1-score} \\
    \midrule
  \multirow{5}*{\rotatebox{90}{\bf Supervision}} & $\text{CWM}_{SVM}$ & 54.5  & 30.6  & 27.8  & 29.1  \\
   & $\text{CWM}_{XGB}$ & 53.7  & 31.0  & 29.5  & 30.2  \\
   & $\text{Fun}_{RF}$ & 57.1  & 29.9  & 27.6  & 28.7  \\
   & $\text{Fun}_{LGBM}$ & 56.0  & 29.5  & 28.8  & 29.1  \\
    & SVACL & 57.4  & \underline{31.2}  & \underline{29.9}  & \underline{30.5}  \\
    \midrule
  \multirow{6}*{\rotatebox{90}{\bf Pre-trained}}  & CodeBERT & 50.9  & 10.6  & 25.0  & 14.9  \\
   & GraphCodeBERT & 53.5  & 12.2  & 25.0  & 16.4  \\
   & CodeT5 & 52.9  & 27.1  & 26.9  & 27.0  \\
   & Codereviewer & 52.6  & 24.3  & 26.5  & 25.4  \\
   & Unixcoder & 53.1  & 12.2  & 25.0  & 16.4  \\
   & RoBERTa & 50.9  & 12.2  & 25.0  & 16.4  \\
    \midrule
  \multirow{6}*{\rotatebox{90}{\bf LLM-based}}  & Qwen2.5-Coder-7b & 56.7  & 28.6  & 28.4  & 28.5  \\
   & Star-Coder2-7b & 57.2  & 28.6  & 27.9  & 28.2  \\
   & Deepseek-Coder-7b & 57.4  & 29.8  & 28.5  & 29.1  \\
   & CodeLlama-7b & \underline{57.8}  & 30.8  & 29.4  & 30.1  \\
    &GPT-3.5 & 55.5  & 27.3  & 27.0  & 27.1  \\
   & DeepSeek-R1 & 56.4  & 27.4  & 29.2  & 28.3  \\
    \midrule
    \textbf{Ours} & \textbf{{\tool}} & \textbf{61.9 } & \textbf{43.5 } & \textbf{32.4 } & \textbf{37.1 } \\
    \bottomrule
    \end{tabular}%
  \label{tab:rq1_1}%
  \vspace{-0.2in}
\end{table}%

\begin{figure}[h]
\vspace{-0.1in}
\centering	
    \includegraphics[width=0.26\textwidth]{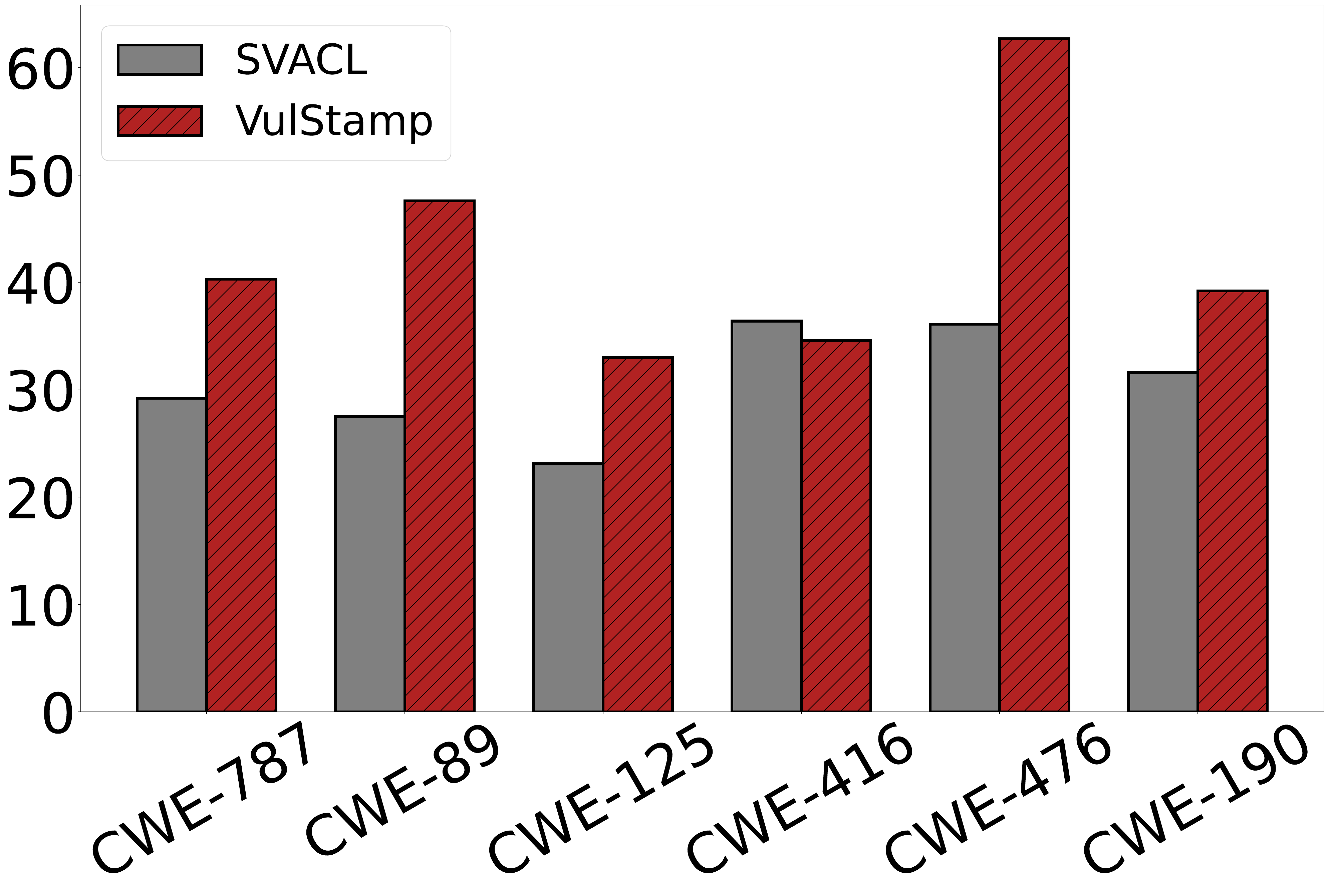}
   \vspace{-0.1in}
    \caption{F1-score of {\tool} and baselines for Top-25 most dangerous CWEs.}
    \label{fig:rq1_2}
 \vspace{-0.15in}
\end{figure}

\textit{\bf {\tool} vs. Supervision-based Methods.}
From  Table~\ref{tab:rq1_1}, we can find that {\tool} significantly outperforms all five baseline methods in terms of all performance metrics. 
Although supervision-based methods can effectively capture the vulnerability code information (i.e., SVACL can achieve the second-best assessment performance), they still greatly suffer from the problems of code noise and lack of code intentions. 
To better understand the effectiveness of {\tool}, we investigated the 25 most dangerous CWEs listed on the CWE website$\footnote{https://cwe.mitre.org/top25/archive/2024/2024\_cwe\_top25.html}$, as they are prevalent in a wide range of applications and systems (e.g., web/desktop applications and operating systems) and should be prioritized for remediation.
Here, to avoid underfitting due to insufficient samples, we omitted vulnerabilities with fewer than 10 samples.  As a result, we compared {\tool} with the best-performing baseline SVACL on six types of vulnerabilities, whose results are shown in Figure~\ref{fig:rq1_2}.
From this figure, we can find that {\tool} always achieves the highest F1-score, showing the superiority of  {\tool} in vulnerability assessment. Note that, since the vulnerability intention reports best fit to address pointer vulnerabilities, {\tool} achieves the best improvement over SVACL for CWE-476 (NULL pointer dereference). 

\textit{\bf {\tool} vs. Pre-trained Model-based Methods.}
From Table~\ref{tab:rq1_1}, we can find that CodeT5 achieves the best results in precision and recall among all six pre-trained model-based methods. However, such results are still far behind those of {\tool}, reflecting that {\tool} captures the vulnerability pattern and the intention of the code more precisely than the pre-training-based baselines.


\textit{\bf {\tool} vs. LLM-based Methods.}
From  Table~\ref{tab:rq1_1}, we can find that CodeLlama-7b shows the best performance in all six LLM-based methods. However, CodeLlama-7b underperforms significantly compared to {\tool} in terms of all four metrics, mainly because it lacks the specific expertise required to identify software vulnerabilities.

\begin{tcolorbox}[width=1.0\linewidth,
    boxsep=1mm,  
    left=0.5mm,    
    right=0.5mm,   
    top=0.05mm,     
    bottom=0.05mm,  
]
\noindent\textbf{Answer to RQ1:} 
Compared with all baselines, {\tool} always performs best on all metrics, showing its superiority in vulnerability assessment.  
\end{tcolorbox}

\subsection{Ablation Study (RQ2)}
The purpose of this experiment is to evaluate how effective each crucial element (such as IDG, VIR, and prompt tuning) is within our {\tool} framework. The findings of the ablation study are summarized in Table~\ref{tab:rq2}, demonstrating that the fully developed VulStamp achieves the highest performance. 



\begin{table}[htbp]
  \vspace{-0.1in}
  \centering
  \caption{Ablation study on {\tool}.}
  \vspace{-0.1in}
  \footnotesize
  \addtolength{\tabcolsep}{-1pt}
    \begin{tabular}{lcccc}
    \toprule
    \rowcolor[rgb]{ .749,  .749,  .749} \textbf{Variants} & \textbf{AUC} & \textbf{Precision} & \textbf{Recall} & \textbf{F1-score} \\
    \midrule
    w/o IDG & 53.7  & 29.6  & 29.2  & 29.4  \\
    w/o IDG-API & 54.4  & 29.9  & 31.4  & 30.6  \\
    w/o IDG-operator & 57.3  & 38.0  & 26.0  & 30.9  \\
    w/o VIR & 53.1  & 28.0  & 27.0  & 27.5  \\
    w/o VIR-exp & 55.6  & 37.9  & 26.6  & 31.3  \\
    w/o VIR-imp & 57.4  & 32.2  & 29.6  & 30.8  \\
    w/o VIR-sco & 54.1  & 28.0  & 27.6  & 27.8  \\
    \midrule
    w/o RL & 55.5  & 30.4  & 28.7  & 29.5  \\
    w/o weighted reward & 59.8  & 34.8  & 27.1  & 30.5  \\
    w/o prompt tuning & 56.8  & 30.7  & 27.8  & 29.2  \\
    w/o suggestion generation & 58.9  & 36.6  & 28.8  & 32.2  \\
    \midrule
    \textbf{{\tool}}  & \textbf{61.9 } & \textbf{43.5 } & \textbf{32.4 } & \textbf{37.1 } \\
    \bottomrule
    \end{tabular}%
  \label{tab:rq2}%
  \vspace{-0.1in}
\end{table}%

\textit{\bf Intention Dependence Graph Construction.}
To demonstrate the effectiveness of the extracted IDG, we performed a comparative study by eliminating the IDG during model training. Furthermore, we sequentially removed the slices' interest points within the IDG (IDG-API and IDG-operator) to assess their individual contributions. Specifically, omitting the IDG led to reductions of 15.3\%, 47.0\%, 11.0\%, and 26.2\% in AUC, precision, recall, and F1-score, respectively. Additionally, removing any interest point from either the API or the operator can reduce the performance of {\tool} by up to 21.2\% in F1-score. This suggests that the source code contains significant redundant information that impairs the model's comprehension of vulnerable code parts.


\textit{\bf Vulnerability Intention Report Generation.}
To assess the impact of incorporating the VIR module into {\tool}, we conducted experiments omitting VIR. The findings reveal that using IDG as input, without incorporating reports on vulnerability-related intentions, results in decreases in AUC, precision, recall, and F1-score by 16.6\%, 55.4\%, 20.0\%, and 34.9\%, respectively. Additionally, omitting either inability, impact, or scope individually (denoted as VIR-exp, VIR-IMP, and VIR-sco) diminishes {\tool}'s performance by up to 14.4\% in AUC and at least 18.5\% in F1-score. This underscores the importance of the VIR components, as well as each individual part, for effectively comprehending vulnerabilities in code.


\textit{\bf Training with Weighted Reward Loss.}
To determine the impact of weighted reward loss on training, we eliminated the aspect of the reward function that targets high-risk vulnerabilities in our experiments. The resulting decline in performance metrics suggests that focusing more on high-risk vulnerabilities enhances the evaluation of the vulnerability function. Specifically, recall decreased by 19.6\%, and the F1-score fell by a significant 21.6\%. We also explored whether incorporating vulnerability fix suggestion generation in a multi-task learning framework could enhance our assessment of vulnerability functions. Omitting the suggestion generation led to an 18.9\% drop in {\tool} precision. These findings demonstrate that emphasizing high-risk vulnerabilities during training more effectively captures vulnerability functions.


\begin{tcolorbox}[width=1.0\linewidth,
    boxsep=1mm,  
    left=0.5mm,    
    right=0.5mm,   
    top=0.05mm,     
    bottom=0.05mm,  
]
\noindent\textbf{Answer to RQ2:} 
Our proposed key components play a crucial role in enhancing the efficiency of {\tool}. By integrating these components, {\tool} is markedly improved for assessing various complex vulnerabilities and generating high-quality repair suggestions.
\end{tcolorbox}


\subsection{Applicability of VulStamp  (RQ3)}
To investigate the applicability of our approach, we conducted experiments on VulStamp with all pre-trained models as shown in Table~\ref{tab:rq1_1}, including CodeBERT~\cite{feng2020codebert}, GraphCodeBERT~\cite{guographcodebert}, RoBERTa~\cite{liu2021robustly}, CodeT5~\cite{wang2021codet5}, and UniXcoder~\cite{guo2022unixcoder}.  

\begin{table}[htbp]
  \vspace{-0.05in}
  \centering
  \caption{Experimental results of components in existing pre-trained model-based methods.}
  \scriptsize
  \vspace{-0.05in}
    \begin{tabular}{ccccc}
    \toprule
    \rowcolor[rgb]{ .749,  .749,  .749} \textbf{Model} & \textbf{AUC} & \textbf{Precision} & \textbf{Recall} & \textbf{F1-score} \\
    \midrule
    \multirow{2}[2]{*}{CodeBERT} & 57.4  & 27.7  & 26.0  & 26.8  \\
          & ($\uparrow$ +12.8\%) & ($\uparrow$ +161.3\%) & ($\uparrow$ +4.0\%) & ($\uparrow$ +79.9\%) \\
    \midrule
    \multirow{2}[2]{*}{GraphCodeBERT} & 58.8  & 27.7  & 27.0  & 27.3  \\
          & ($\uparrow$ +9.9\%) & ($\uparrow$ +127.0\%) & ($\uparrow$ +8.0\%) & ($\uparrow$ +66.5\%) \\
    \midrule
    \multirow{2}[2]{*}{RoBERTa} & 57.0  & 37.4  & 25.8  & 30.5  \\
          & ($\uparrow$ +12.0\%) & ($\uparrow$ +206.6\%) & ($\uparrow$ +3.2\%) & ($\uparrow$ +86.0\%) \\
    \midrule
    \multirow{2}[2]{*}{CodeT5} & 58.7  & 30.2  & 28.7  & 29.4  \\
          & ($\uparrow$ +11.0\%) & ($\uparrow$ +11.4\%) & ($\uparrow$ +6.7\%) & ($\uparrow$ +8.9\%) \\
    \midrule
    \multirow{2}[2]{*}{UniXcoder} & 58.2  & 32.5  & 29.3  & 30.8  \\
          & ($\uparrow$ +9.6\%) & ($\uparrow$ +166.4\%) & ($\uparrow$ +17.2\%) & ($\uparrow$ +87.8\%) \\
    \midrule
    CodeReviewer & 61.9  & 43.5  & 32.4  & 37.1  \\
       (Default)   & ($\uparrow$ +17.7\%) & ($\uparrow$ +79.0\%) & ($\uparrow$ +22.3\%) & ($\uparrow$ +46.1\%) \\
    \bottomrule
    \end{tabular}%
  \label{tab:rq3}%
  \vspace{-0.1in}
\end{table}%

Table~\ref{tab:rq3} shows the experimental results. For example, as shown in Table ~\ref{tab:rq1_1}, the F1-score of the original CodeBERT-based method is 14.9\%, while VulStamp with CodeBERT can achieve an F1-score of 26.8\%, achieving an improvement of 79.9\% on the F1-score.  From this table, we can find that, due to the effectiveness of our proposed techniques (e.g., IDG, VIR, and weighted reward loss), VulStamp can be used to improve the assessment performance of its counterparts that are merely based on pre-trained models. Moreover, we can observe that VulStamp based on CodeReviewer achieves the highest performance, since CodeReviewer itself is designed specifically for code review.


\begin{tcolorbox}[width=1.0\linewidth,
    boxsep=1mm,  
    left=0.5mm,    
    right=0.5mm,   
    top=0.05mm,     
    bottom=0.05mm,  
]
\noindent\textbf{Answer to RQ3:} 
VulStamp is a promising framework for vulnerability assessment that is compatible with a broad range of pre-trained LLMs, enhancing their capabilities in evaluating vulnerabilities. 
\end{tcolorbox}

\subsection{Impacts of Hyper-parameters (RQ4)}
\label{sec:rq4}

\begin{figure}[htpb]
\centering	
    \includegraphics[width=0.45\textwidth]{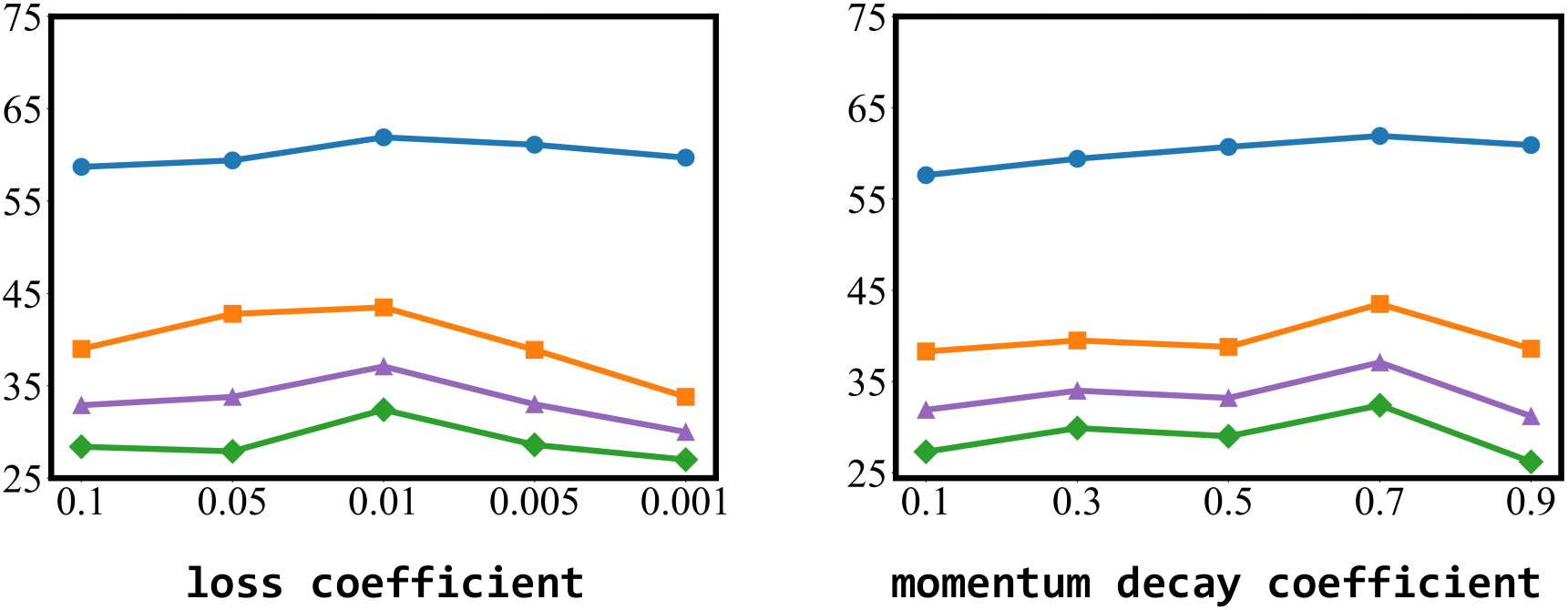}
    \vspace{-0.05in}
    \caption{Performance of {\tool} with different hyper-parameters.}
    \label{fig:rq4}
      \vspace{-0.1in}
\end{figure}

This experiment aims to investigate how hyper-parameters, specifically the weighted reward loss coefficient and the momentum decay coefficient, affect the performance of {\tool}.
Figure~\ref{fig:rq4}
shows the experimental results, where the blue, orange, green, and purple lines indicate the metrics of AUC, precision, recall, and F1-score, respectively.

\textit{\bf Weighted Reward Loss Coefficient.}
As observed in Figure~\ref{fig:rq4}(a), the effectiveness of {\tool} declines when the loss coefficient drops below 0.01, highlighting the critical role of weighted reward loss in enhancing the model evaluation's vulnerability function. Notably, at a loss coefficient of 0.01, {\tool}'s performance peaks, achieving an F1-score of 37.1\%.
Furthermore, when the loss coefficient is below 0.01, the performance of {\tool} tends to gradually decline.

\textit{\bf Momentum Decay Coefficient.}
From Figure~\ref{fig:rq4}(b), we can find that the performance of {\tool}  improves along with the increase of the momentum decay coefficient. As the momentum decay coefficient rises, the reward baseline updates at a slower pace, enhancing its ability to capture the long-term reward trend, minimizing short-term fluctuations, and stabilizing the perception of vulnerability intention.
Upon the momentum decay coefficient reaching 0.7, the AUC, precision, recall, and F1-score achieve their optimum. Beyond this point, all metrics experience a notable decline.
Therefore, we set the momentum decay coefficient to 0.7 during {\tool} training.

\begin{tcolorbox}[width=1.0\linewidth,
    boxsep=1mm,  
    left=0.5mm,    
    right=0.5mm,   
    top=0.05mm,     
    bottom=0.05mm,  
]
\noindent\textbf{Answer to RQ4:} 
The hyper-parameters are crucial in shaping how well {\tool} performs in vulnerability assessment. To enhance this performance, we typically set the weighted reward loss coefficient at 0.01 and the momentum decay coefficient at 0.7. 
\end{tcolorbox}

\section{DISCUSSION}

\subsection{Case Study}

\begin{figure}[htpb]
\centering	
    \includegraphics[width=0.45\textwidth]{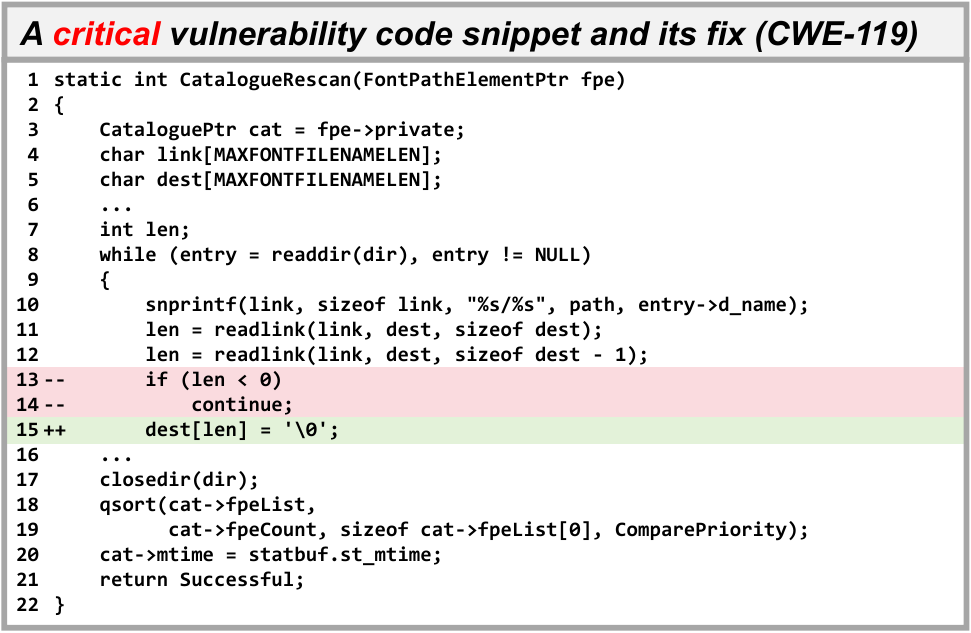}
    \vspace{-0.05in}
    \caption{An example of a vulnerability mis-assessed as medium-risk.}
       \vspace{-0.1in}
    \label{fig:dis1}
\end{figure}

In Figure~\ref{fig:dis1}, SVACL incorrectly classified CWE-119 as a medium-risk vulnerability in the evaluation dataset, whereas {\tool} correctly identified it as a critical-risk issue.
The \texttt{readlink} function is utilized to fetch the contents of a symbolic link, essentially retrieving the path of the target file that the symbolic link references. The third parameter indicates the buffer size designated for storing the target path. In this context, the buffer size is initialized to \texttt{sizeof} \texttt{dest}. Nonetheless, if the target path, including the null terminator $\setminus 0$, surpasses the \texttt{dest} buffer's capacity, the \texttt{readlink} function will extend writing operations beyond the buffer size in \texttt{dest}, leading to a buffer overflow vulnerability. The risk manifests under a specific directory configuration, requiring the presence of symbolic link files in the directory, with the target path's length marginally exceeding the buffer's limit. SVACL struggles to fully capture the vulnerability's implications, which can cause it to overlook potential threats, resulting in false positives.

{\tool} comprehends the significance and extent of the vulnerability, enabling it to concentrate on its purpose. Additionally, it can monitor the specific local code details causing the vulnerability, allowing for an accurate evaluation of its classification as a critical-risk.

\subsection{Impact of  Intention Reports Generated by LLMs}

To evaluate the effectiveness of generating VIRs, we conducted experiments using different LLMs.
The findings presented in~\tablename~\ref{tab:dis2} suggest that both GPT-3.5-turbo and DeepSeek-R1 produce VIRs, enabling {\tool} to outperform all baseline models.
The VIRs produced by GPT-3.5-turbo enhanced {\tool} by 7.8\%, 39.4\%, 8.4\%, and 21.6\% in AUC, precision, recall, and F1-score, respectively, over the top baseline. Similarly, the {\tool} used in conjunction with DeekSeek-R1 showed improvements of 8.4\%, 19.2\%, 4.3\%, and 11.1\% for AUC, precision, recall, and F1-score, respectively, when compared to the best baseline.
This suggests that while the VIRs produced by DeekSeek-R1 based on instructions are effective, certain constraints remain when compared to GPT-3.5-turbo. These results indicate that the VIRs generated by different LLMs can all enhance the model's comprehension of software vulnerability intentions, thus improving the effectiveness of the evaluation.

\begin{table}[htbp]
\vspace{-0.1in}
  \centering
  \caption{Comparison for VIR generated by different LLMs.}
  \vspace{-0.05in}
  \footnotesize
    \begin{tabular}{lcccc}
    \toprule
    \rowcolor[rgb]{ .749,  .749,  .749} \textbf{Model} & \textbf{AUC} & \textbf{Precision} & \textbf{Recall} & \textbf{F1-score} \\
    \midrule
    GPT-3.5 & 61.9  & \textbf{43.5}  & \textbf{32.4}  & \textbf{37.1}  \\
    DeepSeek-R1 & \textbf{62.2}  & 37.2  & 31.2  & 33.9 \\
    \bottomrule
    \end{tabular}%
  \label{tab:dis2}%
  \vspace{-0.1in}
\end{table}%



\subsection{Quality of Generated Repair Suggestions}


The results of the ablation study demonstrate the efficacy of suggestion generation. To further investigate the performance of the generated suggestions, we conducted a comparison with the latest baseline, VulAdvisor, which is designed to address vulnerabilities. We specifically used the performance metrics BLEU-4, ROUGE-L, METEOR, and BERTScore, as employed by VulAdvisor. Table~\ref{tab:dis4} presents the results of our experiments. We found that the suggestions produced by {\tool} surpassed those of VulAdvisor on all four metrics. Notably, in terms of BLEU-4, it outperformed by 3.6\%. The suggestions generated by {\tool} align more closely with the reference suggestions in both semantics and grammatical structure. Furthermore, they exhibit improved accuracy in vocabulary usage and phrasing, thus more effectively matching the intent conveyed in real bug repair suggestions.

\begin{table}[htbp]
\vspace{-0.1in}
  \centering
  \caption{Comparison for vulnerability repair suggestions.}
  \vspace{-0.05in}
  \footnotesize
    \begin{tabular}{lcccc}
    \toprule
    \rowcolor[rgb]{ .749,  .749,  .749} \textbf{Model} & \multicolumn{1}{l}{\textbf{BLEU}} & \multicolumn{1}{l}{\textbf{ROUGE-L}} & \multicolumn{1}{l}{\textbf{METEOR}} & \multicolumn{1}{l}{\textbf{BERTScore}} \\
    \midrule
    VulAdvisor & 41.2  & 58.4  & 33.1  & 83.6  \\
        {\tool}  & \textbf{42.7}  & \textbf{58.7}  & \textbf{33.7}  & \textbf{84.1}  \\
    \bottomrule
    \end{tabular}%
  \label{tab:dis4}%
  \vspace{-0.1in}
\end{table}%

\section{RELATED WORK}

\subsection{Descriptive Feature-based Vulnerability Assessment}

During the application process, a vulnerability description is provided by the CVE-ID applicant, detailing the vulnerability in natural language. This description typically encompasses essential information, technical specifics, the extent of impact, and potential repercussions of the vulnerability.

Various methods have been investigated to demonstrate the effectiveness of vulnerability descriptions for software vulnerability assessment.
For example, Han et al.~\cite{han2017learning} reframed the challenge of assessing vulnerabilities as a text classification issue, employed the continuous Skip-Gram model to train word embeddings from a gathered corpus of vulnerability descriptions, and developed a single-layer shallow CNN to extract sentence-level characteristics from vulnerability descriptions.
Sahin et al.~\cite{sahin2019conceptual} adopted the approach of Han et al.~\cite{han2017learning}, utilizing word vectors for extracting features and employing convolutional neural networks to construct a predictive model.
Le et al.~\cite{le2019automated} introduced a structured method that integrates both character and word features to automatically conduct software vulnerability evaluations considering concept drift, utilizing software vulnerability descriptions. Their approach employs a specially designed time-based cross-validation technique to identify the optimal model for each vulnerability characteristic, chosen from eight distinct natural language processing representations and six established machine learning models.
Le et al.~\cite{le2022survey} conducted a review of earlier research activities, emphasizing optimal practices for assessing and ranking software vulnerabilities driven by data.
Sun et al.~\cite{sun2023automatic} employed BERT-MRC to isolate vulnerability components from their descriptions and used these elements throughout the descriptions to assess six different metrics.

The descriptive features of vulnerabilities are very useful for an initial understanding of vulnerabilities and their potential threats.
However, the actual occurrence and exploitation of vulnerabilities are often closely related to the specific implementation of the code, and the code features can provide more detailed and low-level technical details.
In addition, the vulnerability description is more dependent on the applicant's understanding, and the writing style and understanding of the description obtained may not be complete.

\subsection{Code Feature-based Vulnerability Assessment}

In recent times, there has been an increase in the use of methods for assessing vulnerabilities by analyzing susceptible code.
For example, Le et al.~\cite{le2022use} explored how vulnerable statements could be used to create evaluation models, integrating the context of these statements, resulting in an improvement in performance of 8.9\%.
Hao et al.~\cite{hao2023novel} introduced a method to assess vulnerability severity, which integrates both the call graph of the function and the vulnerability attribute graph. This approach employs the vulnerability attribute graph to depict the vulnerability's severity based on the code's semantics, and utilizes a graph attention neural network to enhance the accuracy of the vulnerability severity evaluation.
Xue et al.~\cite{xue2025towards} integrated confidence-based replay techniques with regularization strategies for continuous learning by employing source code and vulnerability descriptions alongside the pre-trained CodeT5 model to develop a hybrid prompt. 


Compared with existing methods, in this paper, we introduce a novel intention-guided vulnerability assessment based on LLMs. 
Unlike previous studies, we propose to extract vulnerability intention statements from the code and analyze the exploitability, impact, and scope of the code through LLMs to obtain the vulnerability intentions.
In our approach, more attention is paid to high-risk vulnerabilities to prevent them from being misjudged as low-risk to avoid major losses. 
Last but not least, our approach supports the generation of high-quality suggestions for vulnerability repair. 
To the best of our knowledge, our work is the first attempt to combine the merits of intention-oriented syntactic code characteristics and the semantical natural language processing capabilities of LLMs, thus enhancing the comprehension of vulnerabilities.



\section{CONCLUSION}

This paper introduces {\tool}, a novel LLM-based framework that enables a precise assessment of software vulnerabilities and provides constructive repair suggestions. 
With our proposed intention-guided data processing method and designed prompt template,  {\tool} can not only extract syntactical information for the harmful intention of identified vulnerabilities, but also produce severity reports for these vulnerabilities, including their exploitability, impact, and scope. 
By prompt-tuning on a pre-trained LLM model using the collected intention-oriented information, our approach forms a code reviewer model for both vulnerability assessment and repair goals. 
During the prompt-tuning, to alleviate the problem of imbalanced data associated with vulnerability types, we employed an efficient gradient-enhanced model training scheme based on reinforcement learning, which can significantly improve the accuracy of the assessment and the quality of repair suggestions. 
Comprehensive experimental results on well-known LLMs and vulnerability benchmarks demonstrate the effectiveness of {\tool}.


%




\newpage

\bibliographystyle{unsrt}
\balance
\bibliography{sn-bibliography.bib}

\begin{thebibliography}{10}

\bibitem{cao2024snopy}
Sicong Cao, Xiaobing Sun, Xiaoxue Wu, David Lo, Lili Bo, Bin Li, Xiaolei Liu, Xingwei Lin, and Wei Liu.
\newblock Snopy: Bridging sample denoising with causal graph learning for effective vulnerability detection.
\newblock In {\em Proceedings of the International Conference on Automated Software Engineering (ASE)}, pages 606--618, 2024.

\bibitem{zhao2024coding}
Yu~Zhao, Lina Gong, Zhiqiu Huang, Yongwei Wang, Mingqiang Wei, and Fei Wu.
\newblock Coding-ptms: How to find optimal code pre-trained models for code embedding in vulnerability detection?
\newblock In {\em Proceedings of the International Conference on Automated Software Engineering (ASE)}, pages 1732--1744, 2024.

\bibitem{zhang2024vuladvisor}
Jian Zhang, Chong Wang, Anran Li, Wenhan Wang, Tianlin Li, and Yang Liu.
\newblock Vuladvisor: Natural language suggestion generation for software vulnerability repair.
\newblock In {\em Proceedings of the International Conference on Automated Software Engineering (ASE)}, pages 1932--1944, 2024.

\bibitem{antunes2010vulnerability}
Joao Antunes, Nuno Neves, Miguel Correia, Paulo Verissimo, and Rui Neves.
\newblock Vulnerability discovery with attack injection.
\newblock {\em IEEE Transactions on Software Engineering}, 36(3):357--370, 2010.

\bibitem{zuo2019does}
Chaoshun Zuo, Zhiqiang Lin, and Yinqian Zhang.
\newblock Why does your data leak? uncovering the data leakage in cloud from mobile apps.
\newblock In {\em Proceedings of the IEEE Symposium on Security and Privacy (SP)}, pages 1296--1310, 2019.

\bibitem{croft2023data}
Roland Croft, M~Ali Babar, and M~Mehdi Kholoosi.
\newblock Data quality for software vulnerability datasets.
\newblock In {\em Proceedings of the International Conference on Software Engineering (ICSE)}, pages 121--133, 2023.

\bibitem{dissanayake2022empirical}
Nesara Dissanayake, Asangi Jayatilaka, Mansooreh Zahedi, and Muhammad~Ali Babar.
\newblock An empirical study of automation in software security patch management.
\newblock In {\em Proceedings of the International Conference on Automated Software Engineering (ASE)}, pages 1--13, 2022.

\bibitem{balsam2023automated}
Artur Balsam, Micha{\l} Walkowski, Maciej Nowak, Jacek Oko, and S{\l}awomir Sujecki.
\newblock Automated calculation of cvss v3. 1 temporal score based on apache log4j 2021 vulnerabilities.
\newblock In {\em Proceedings of the International Conference on Software, Telecommunications and Computer Networks (SoftCOM)}, pages 1--3, 2023.

\bibitem{csw2023ransomware}
{Cyber Security Works}, {Ivanti}, {Cyware}, and {Securin}.
\newblock Ransomware spotlight report 2023.
\newblock Technical report, Cyber Security Works, 2023.
\newblock Accessed: 2025-05-30.

\bibitem{CVE-2021-45046}
CVE.
\newblock \url{https://nvd.nist.gov/vuln/detail/CVE-2021-45046}.
\newblock 2021.

\bibitem{megavul}
Chao Ni, Liyu Shen, Xiaohu Yang, Yan Zhu, and Shaohua Wang.
\newblock Megavul: A c/c++ vulnerability dataset with comprehensive code representations.
\newblock In {\em Proceedings of International Conference on Mining Software Repositories (MSR)}, pages 738--742, 2024.

\bibitem{sun2023automatic}
Xiaobing Sun, Zhenlei Ye, Lili Bo, Xiaoxue Wu, Ying Wei, Tao Zhang, and Bin Li.
\newblock Automatic software vulnerability assessment by extracting vulnerability elements.
\newblock {\em Journal of Systems and Software}, 204:111790, 2023.

\bibitem{xue2025towards}
Jiacheng Xue, Xiang Chen, Jiyu Wang, and Zhanqi Cui.
\newblock Towards prompt tuning-based software vulnerability assessment with continual learning.
\newblock {\em Computers \& Security}, 150:104184, 2025.

\bibitem{han2017learning}
Zhuobing Han, Xiaohong Li, Zhenchang Xing, Hongtao Liu, and Zhiyong Feng.
\newblock Learning to predict severity of software vulnerability using only vulnerability description.
\newblock In {\em Proceedings of International Conference on Software Maintenance and Evolution (ICSME)}, pages 125--136, 2017.

\bibitem{sahin2019conceptual}
Sefa~Eren Sahin and Ayse Tosun.
\newblock A conceptual replication on predicting the severity of software vulnerabilities.
\newblock In {\em Proceedings of International Conference on Evaluation and Assessment in Software Engineering}, pages 244--250, 2019.

\bibitem{le2019automated}
Triet Huynh~Minh Le, Bushra Sabir, and Muhammad~Ali Babar.
\newblock Automated software vulnerability assessment with concept drift.
\newblock In {\em Proceedings of International Conference on Mining Software Repositories (MSR)}, pages 371--382, 2019.

\bibitem{le2022use}
Triet Huynh~Minh Le and M~Ali Babar.
\newblock On the use of fine-grained vulnerable code statements for software vulnerability assessment models.
\newblock In {\em Proceedings of International Conference on Mining Software Repositories (MSR)}, pages 621--633, 2022.

\bibitem{hao2023novel}
Jingwei Hao, Senlin Luo, and Limin Pan.
\newblock A novel vulnerability severity assessment method for source code based on a graph neural network.
\newblock {\em Information and Software Technology}, 161:107247, 2023.

\bibitem{dang2015comparing}
Duy Dang-Pham and Siddhi Pittayachawan.
\newblock Comparing intention to avoid malware across contexts in a byod-enabled australian university: A protection motivation theory approach.
\newblock {\em Computers \& Security}, 48:281--297, 2015.

\bibitem{iannone2022secret}
Emanuele Iannone, Roberta Guadagni, Filomena Ferrucci, Andrea De~Lucia, and Fabio Palomba.
\newblock The secret life of software vulnerabilities: A large-scale empirical study.
\newblock {\em IEEE Transactions on Software Engineering}, 49(1):44--63, 2022.

\bibitem{davis2004processes}
Noopur Davis, Watts Humphrey, Samuel~T Redwine, Gerlinde Zibulski, and Gary McGraw.
\newblock Processes for producing secure software.
\newblock {\em IEEE Security \& Privacy}, 2(3):18--25, 2004.

\bibitem{shahzad2012large}
Muhammad Shahzad, Muhammad~Zubair Shafiq, and Alex~X Liu.
\newblock A large scale exploratory analysis of software vulnerability life cycles.
\newblock In {\em 2012 34th International Conference on Software Engineering (ICSE)}, pages 771--781. IEEE, 2012.

\bibitem{sabottke2015vulnerability}
Carl Sabottke, Octavian Suciu, and Tudor Dumitraș.
\newblock Vulnerability disclosure in the age of social media: Exploiting twitter for predicting $\{$Real-World$\}$ exploits.
\newblock In {\em 24th USENIX security symposium (USENIX security 15)}, pages 1041--1056, 2015.

\bibitem{chen2025chatgpt}
Chong Chen, Jianzhong Su, Jiachi Chen, Yanlin Wang, Tingting Bi, Jianxing Yu, Yanli Wang, Xingwei Lin, Ting Chen, and Zibin Zheng.
\newblock When chatgpt meets smart contract vulnerability detection: How far are we?
\newblock {\em ACM Transactions on Software Engineering and Methodology}, 34(4):1--30, 2025.

\bibitem{yan2024stealing}
Kailun Yan, Xiaokuan Zhang, and Wenrui Diao.
\newblock Stealing trust: Unraveling blind message attacks in web3 authentication.
\newblock In {\em Proceedings of the 2024 on ACM SIGSAC Conference on Computer and Communications Security}, pages 555--569, 2024.

\bibitem{wen2024livable}
Xin-Cheng Wen, Cuiyun Gao, Feng Luo, Haoyu Wang, Ge~Li, and Qing Liao.
\newblock Livable: exploring long-tailed classification of software vulnerability types.
\newblock {\em IEEE Transactions on Software Engineering}, 2024.

\bibitem{cvss1}
Peter Mell, Karen Scarfone, and Sasha Romanosky.
\newblock A complete guide to the common vulnerability scoring system version 2.0.
\newblock {\em FIRST-forum of incident response and security teams}, 1, 2007.

\bibitem{cvss}
Peter Mell, Karen Scarfone, and Sasha Romanosky.
\newblock Common vulnerability scoring system.
\newblock {\em IEEE Security \& Privacy}, 4(6):85--89, 2006.

\bibitem{cvss3}
Pontus Johnson, Robert Lagerstr\"{o}m, Mathias Ekstedt, and Ulrik Franke.
\newblock Can the common vulnerability scoring system be trusted? a bayesian analysis.
\newblock {\em IEEE Transactions on Dependable and Secure Computing}, 15(6):1002--1015, 2016.

\bibitem{cvssicse}
Shengyi Pan, Lingfeng Bao, Jiayuan Zhou, Xing Hu, Xin Xia, and Shanping Li.
\newblock Towards more practical automation of vulnerability assessment.
\newblock In {\em Proceedings of International Conference on Software Engineering (ICSE)}, pages 148:1--148:13, 2024.

\bibitem{CVE-2023-34152}
CVE.
\newblock \url{https://www.cve.org/CVERecord?id=CVE-2023-34152}.
\newblock 2023.

\bibitem{yin2024multitask}
Xin Yin, Chao Ni, and Shaohua Wang.
\newblock Multitask-based evaluation of open-source llm on software vulnerability.
\newblock {\em IEEE Transactions on Software Engineering}, 2024.

\bibitem{wang2024contracttinker}
Che Wang, Jiashuo Zhang, Jianbo Gao, Libin Xia, Zhi Guan, and Zhong Chen.
\newblock Contracttinker: Llm-empowered vulnerability repair for real-world smart contracts.
\newblock In {\em Proceedings of the 39th IEEE/ACM International Conference on Automated Software Engineering}, pages 2350--2353, 2024.

\bibitem{cheng2021deepwukong}
Xiao Cheng, Haoyu Wang, Jiayi Hua, Guoai Xu, and Yulei Sui.
\newblock Deepwukong: Statically detecting software vulnerabilities using deep graph neural network.
\newblock {\em ACM Transactions on Software Engineering and Methodology (TOSEM)}, 30(3):1--33, 2021.

\bibitem{yadavally2024learning}
Aashish Yadavally, Yi~Li, Shaohua Wang, and Tien~N Nguyen.
\newblock A learning-based approach to static program slicing.
\newblock {\em Proceedings of the ACM on Programming Languages}, 8(OOPSLA1):83--109, 2024.

\bibitem{thome2018integrated}
Julian Thome, Lwin~Khin Shar, Domenico Bianculli, and Lionel Briand.
\newblock An integrated approach for effective injection vulnerability analysis of web applications through security slicing and hybrid constraint solving.
\newblock {\em IEEE Transactions on Software Engineering}, 46(2):163--195, 2018.

\bibitem{cai2025adapting}
Xuemeng Cai and Lingxiao Jiang.
\newblock Adapting knowledge prompt tuning for enhanced automated program repair.
\newblock In {\em 2025 IEEE International Conference on Software Analysis, Evolution and Reengineering (SANER)}, pages 360--371. IEEE, 2025.

\bibitem{feng2020codebert}
Zhangyin Feng, Daya Guo, Duyu Tang, Nan Duan, Xiaocheng Feng, Ming Gong, Linjun Shou, Bing Qin, Ting Liu, Daxin Jiang, et~al.
\newblock Codebert: A pre-trained model for programming and natural languages.
\newblock In {\em Proceedings of Empirical Methods in Natural Language Processing (EMNLP)}, pages 1536--1547, 2020.

\bibitem{guographcodebert}
Daya Guo, Shuo Ren, Shuai Lu, Zhangyin Feng, Duyu Tang, Liu Shujie, Long Zhou, Nan Duan, Alexey Svyatkovskiy, Shengyu Fu, et~al.
\newblock Graphcodebert: Pre-training code representations with data flow.
\newblock In {\em Proceedings of International Conference on Learning Representations (ICLR)}, 2021.

\bibitem{liu2021robustly}
Zhuang Liu, Wayne Lin, Ya~Shi, and Jun Zhao.
\newblock A robustly optimized bert pre-training approach with post-training.
\newblock In {\em China national conference on Chinese computational linguistics}, pages 471--484. Springer, 2021.

\bibitem{li2022automating}
Zhiyu Li, Shuai Lu, Daya Guo, Nan Duan, Shailesh Jannu, Grant Jenks, Deep Majumder, Jared Green, Alexey Svyatkovskiy, Shengyu Fu, et~al.
\newblock Automating code review activities by large-scale pre-training.
\newblock In {\em Proceedings of ACM Joint European Software Engineering Conference and Symposium on the Foundations of Software Engineering (FSE)}, pages 1035--1047, 2022.

\bibitem{guo2022unixcoder}
Daya Guo, Shuai Lu, Nan Duan, Yanlin Wang, Ming Zhou, and Jian Yin.
\newblock Unixcoder: Unified cross-modal pre-training for code representation.
\newblock In {\em Proceedings of Annual Meeting of the Association for Computational Linguistics (ACL)}, pages 7212--7225, 2022.

\bibitem{wang2021codet5}
Yue Wang, Weishi Wang, Shafiq Joty, and Steven~CH Hoi.
\newblock Codet5: Identifier-aware unified pre-trained encoder-decoder models for code understanding and generation.
\newblock In {\em Proceedings of Empirical Methods in Natural Language Processing (EMNLP)}, pages 8696--8708, 2021.

\bibitem{devign}
Yaqin Zhou, Shangqing Liu, Jingkai Siow, Xiaoning Du, and Yang Liu.
\newblock Devign: Effective vulnerability identification by learning comprehensive program semantics via graph neural networks.
\newblock In {\em Proceedings of Advances in Neural Information Processing Systems (NeurIPS)}, pages 10197--10207, 2019.

\bibitem{bigvul}
Jiahao Fan, Yi~Li, Shaohua Wang, and Tien~N. Nguyen.
\newblock A c/c++ code vulnerability dataset with code changes and cve summaries.
\newblock In {\em Proceedings of International Conference on Mining Software Repositories (MSR)}, pages 508--512, 2020.

\bibitem{le2022survey}
Triet~HM Le, Huaming Chen, and M~Ali Babar.
\newblock A survey on data-driven software vulnerability assessment and prioritization.
\newblock {\em ACM Computing Surveys}, 55(5):1--39, 2022.

\end{thebibliography}

\end{document}